\documentclass[a4paper,11pt]{article}
\pdfoutput=1 % if your are submitting a pdflatex (i.e. if you have
\usepackage{amssymb,amsmath}

\usepackage{jcappub} % for details on the use of the package, please
\usepackage[T1]{fontenc} % if needed
\usepackage{newtxtext,newtxmath}
\usepackage{ae,aecompl}
\usepackage{hyperref}
\usepackage{cleveref}
\usepackage{changes}
\usepackage{enumitem}
\usepackage{xspace}
\usepackage[normalem]{ulem}

\def\prl{Physical Review Letters}
\def\prd{Physical Review D}

\def\prc{Physical Review C}

\def\jcap{JCAP}

\def\apj{ApJ}
\def\apjl{ApJL}

\def\apss{Astroph. Sp. Sci.}
\def\aap{A\&A}

\def\araa{ARAA}
\def\jgr{Journal of Geophys. Res.}
\def\mnras{MNRAS}

\def\physrep{Phys. Rep.}
\def\ssr{Space Sci. Reviews}

\def\nphysa{Nucl. Phys. A}

\def\kms{\rm ~km~s^{-1}}

\def\lsim{\;\raise0.3ex\hbox{$<$\kern-0.75em\raise-1.1ex\hbox{$\sim$}}\;}
\def\gsim{\;\raise0.3ex\hbox{$>$\kern-0.75em\raise-1.1ex\hbox{$\sim$}}\;}

\def\beq#1{\begin{equation}\label{#1}}
\def\eeq{\end{equation}}
\def\beqa#1{\begin{eqnarray}\label{#1}}
\def\eeqa{\end{eqnarray}}

\def\myfrac#1#2{\left(\frac{#1}{#2}\right)}

\def\comment#1{\relax}

\newcommand{\ergs}{\ensuremath{\mathrm{erg\,s}^{-1}}\xspace}

\newcommand{\be}{\begin{eqnarray}}
\newcommand{\ee}{\end{eqnarray}}

\title{\boldmath Antistars as possible sources of antihelium cosmic rays}

\author[a] {A.M.~Bykov,}
\author[b,1] {K.A.~Postnov
\note{Corresponding author.}}
\author[c,d] {A.E.~Bondar,}
\author[e,b] {S.I.~Blinnikov,}
\author[d,f] {A.D.~Dolgov,}

\affiliation[a]{Ioffe Institute, Politechnicheskaya 26, St Petersburg}
\affiliation[b]{Sternberg Astronomical Institute, M.V. Lomonosov Moscow State University,\\ 13, Universitetskij pr., 119234, Moscow, Russia}
\affiliation[c]{Budker INP, Lavrentieva 11, Novosibirsk, 630090, Russia}
\affiliation[d]{Department of Physics, Novosibirsk State University, \\Pirogova 2, 630090, Novosibirsk, Russia}
\affiliation[e]{NRC ``Kurchatov institute'' - Kurchatov Sq. 1, 123182, Moscow,  Russia }
\affiliation[f]{{Bogolyubov Laboratory of Theoretical Physics, Joint Institute for Nuclear Research,
Joliot-Curie st. 6, Dubna, Moscow region, 141980 Russia}}

\emailAdd{byk@astro.ioffe.ru}
\emailAdd{pk@sai.msu.ru}
\emailAdd{A.E.Bondar@inp.nsk.su}
\emailAdd{sergei.blinnikov@itep.ru}
\emailAdd{dolgov@nsu.ru}

\abstract{A minor population of antistars in galaxies has been predicted by some of non-standard models of baryogenesis and
nucleosynthesis in the early Universe, and their presence is not yet excluded by the currently available observations. Detection of an unusually high abundance of antinuclei in cosmic rays can probe the baryogenesis scenarios in the early Universe.
Recent report of the \textit{AMS-02} collaboration on the tentative detection of a few antihelium nuclei in GeV cosmic rays provided a great hope on the progress in this issue. We discuss possible sources of antinuclei in cosmic rays from antistars which are predicted in a modified Affleck-Dine baryogenesis scenario by Dolgov and Silk (1993). 
The model allows us to estimate the expected fluxes and isotopic content of 
antinuclei in the GeV cosmic rays produced in scenarios involving antistars. We show that 
the flux of antihelium CRs reported by the \textit{AMS-02} experiment can be explained by Galactic anti-nova outbursts, thermonuclear anti-SN Ia explosions, a collection of flaring antistars or an extragalactic source with abundances not violating existing gamma-ray and microlensing constraints on the antistar population. 
}

\begin{document}

\maketitle
\flushbottom

%\tcred{BELOW IS THE OLD TEXT!!! FEEL FREE TO ADD ANY SECTIONS}

\section{Introduction}

The problem of antimatter in the Universe remains topical for many years (see, e.g. \cite{1976ARA&A..14..339S} for a review). The antimatter content in the Universe  
can be expressed through the baryon-to-photon ratio, 
$\eta = (N_B - N_{\bar B} ) / N_\gamma$. The observed average value 
over the entire universe is $\eta\approx 6 \cdot 10^{-10}$. This small value (aka baryon asymmetry problem) has not been explained and is a matter of intensive investigations 
\cite{2012NJPh...14i5012C}.
However, negative $\eta$ in some regions of the 
universe are not excluded  and would indicate 
these regions are antimatter-dominated.

Both direct and indirect searches for antimatter in space is ongoing for many decades (see e.g. \cite{1976ARA&A..14..339S,1982ApJ...260...20A,1983Ap&SS..96..171S,Ormes1997ApJ...482L.187O,2000Belotsky, 2002RPPh...65.1243G,2014HyInt.228..100V,AMS_pbar16,AMS_19}).  
The interest to antimatter has been revived by a recent analysis of unidentified gamma-ray sources within \textit{Fermi}/LAT gamma-ray error boxes \cite{2021PhRvD.103h3016D} which can be associated with Galactic compact objects (stars) composed of antimatter (antistars) \cite{2022JCAP...03..009B}. Antistars can appear in some physical scenarios of cosmic baryogenesis leading to the formation of areas in the early Universe (at a temperature of the QCD phase transition of $T\sim 100$~MeV) with a non-standard high baryon asymmetry while keeping the overall small baryon asymmetry at the observed level 
${\eta} \approx 6 \cdot 10^{-10}$ \cite{1993PhRvD..47.4244D,2009NuPhB.807..229D}.

Even more intriguing have been recent reports on the possible detection of antihelium nuclei in low-energy cosmic rays in the \textit{AMS-02} experiment (e.g. \cite{2022cosp...44.3071T,2022cosp...44.2083C}). If confirmed, these measurements could be a strong indication of a non-standard production of antihelium in the Galaxy or the presence of extragalactic antimatter CR sources 
(e.g., by the physics beyond the standard model \cite{arxiv.2211.00025}).   

In this paper, we consider possible explanations of the tentative detection of antihelium  nuclei by a population of Galactic or extragalactic antistars. 
Astrophysical constraints on the properties of possible  Galactic antistar population have been discussed earlier \cite{2007NuPhB.784..132B,2014JETPL..98..519D,2014PhRvD..89b1301D,2014HyInt.228..100V,2015PhRvD..92b3516B,sym14091953,2022JCAP...03..009B}. In Section \ref{s-BBN} we briefly discuss a non-standard nucleosynthesis in the early Universe at large baryon-to-photon ratios $|\eta|\gg 10^{-9}$. In Section \ref{s:AMS} we summarize the data on antihelium nuclei tentatively detected by the \textit{AMS-02} experiment and discuss their propagation in the Galaxy. In Section \ref{s:antiCR} we propose interpretation of the \textit{AMS-02} results by the presence of antistars in the Galaxy. Our conclusions are presented in  Section \ref{s:conclusion}.

\section{Anti-BBN \label{s-BBN}}

The outcome of the primordial nucleosynthesis in the regions filled with antimatter should be drastically different from the
standard Big-Bang nucleosynthesis (BBN). There are at least two reasons for that. First, the standard BBN proceeds with a very low baryon-to-photon
ratio $\eta  \approx 10^{-9}$ 
\footnote{When BBN is considered, the notation $\eta$ is used
for the baryon-to-photon ratio, while in the case of baryogenesis the same quantity is denoted as $\beta$}. 
According to the scenario of Refs.~\cite{1993PhRvD..47.4244D,2009NuPhB.807..229D} based on the modified Affleck-Dine baryogenesis \cite{1985NuPhB.249..361A}, the  primordial nucleosynthesis
in the antimatter condensations (`anti-nucleosynthesis')  proceeds in
the bubbles with an extremely high density of baryons 
(we abbreviate them as HBB = high baryon bubbles) 
where $|\eta|$ may be close to unity.

The second reason for the difference between the standard 
BBN and the BBN inside the bubbles with high $\eta$ is that the  
laws of temperature and density evolution inside the 
bubbles with high (anti)matter density strongly differ from those inherent to the standard BBN. 
The expansion of such bubbles happened to be  much slower than the standard cosmological
expansion due to a large density contrast with respect 
to the cosmological background. For example, if the (anti)baryonic density
of bubbles were high enough, they would completely decouple from the overall 
cosmological expansion and formed black holes. Bubbles with smaller 
mass, due to smaller, but still large,  $|\eta|$, while being not completely 
decoupled from the Hubble expansion, 
still have considerable density excess that leads to
significantly diminished expansion and cooling rates.

BBN with a very high value of $\eta$ was studied in Refs.~\cite{2004PThPh.112..971M,2005PhRvD..72l3505M}. No significant difference with respect to the
standard BBN was found for abundances of $^4$He, $^3$He, and deuterium. But this could be expected, because the value
of $\eta$ adopted in these papers was smaller than $10^{-2}$. BBN with $\eta$ an order of magnitude larger, $\eta\sim 0.1$, was studied in paper \cite{2020PhRvD.102b3503A} with
essentially the same results for abundances of light elements.
This is also a naturally expected result because 
a considerable difference may be reached only for 
$\eta \sim 1$.
For such a large $\eta$  
the conditions inside HBBs resemble those inside stars in the sense that the ratio of the baryon to photon densities is quite large and the temperature inside HBB is close 
to that inside dense purely hydrogen stars. 
However, in the modified Affleck-Dine baryogenesis model \cite{1993PhRvD..47.4244D,2009NuPhB.807..229D} the value of $\eta$ should be somewhat 
below unity due to dilution of the baryon number density 
created by the `rotation' of the Affleck-Dine scalar baryon, by
the entropy released from the decay of the inflaton field. 
Note that this model predicts a log-normal distribution of HBB masses which can lead to a primordial black hole formation with log-normal mass distribution at the QCD phase transition \cite{2016JCAP...11..036B,2020JCAP...12..017D}.
To fulfill the
job of making primordial black holes with a log-normal mass
spectrum, the sufficient value of $\eta$ could be about $\eta \sim T_{\rm QCD}/m_p\sim 0.1$, because HBBs with such $\eta$ would have density
contrast of order unity at the QCD phase transition.

However, as we have already mentioned, the nucleosynthesis calculations inside HBBs have been done
by assuming the standard expanding cosmological background, while 
the temperature inside the bubbles drops down much slower,  which might lead to the formation of heavier species up to 
%Fe or Ni.
the iron group elements.
Therefore, one may expect the existence and observation of even anti-iron primordial nuclei originated in such bubbles.

The slow variation of temperature and density inside 
HBBs also makes the nucleosynthesis inside them similar to the stellar one and
%effect \tcviol{already mentioned above that}
could strongly change the outcome of the 'standard' BBN,
much stronger than a high value of $\eta$.
%is a slow change of temperature and density inside HBBs because %such bubbles might essentially decouple from the cosmological 
%expansion. This phenomenon
In particular, for a certain, almost fixed, value of temperature
the light element abundances could be strongly modified
%might have a strong impact on the abundances of light elements, 
as illustrated in Figure 3  from Ref.~\cite{1967ApJ...148....3W}.
One can see, for example, that at some temperature $^4$He is produced in equal amounts to deuterium.
%\begin{figure}[htbp]
%\begin{center}
%\includegraphics[scale=0.45,angle=-90]{BBN-of-time.pdf}
%\caption{}
%\label{JWST-1}
%\end{center}
%\end{figure}

To conclude, there  do not yet exist rigorous calculations of light (and heavy?) element production inside HBBs. Still it seems
possible that the abundances of helium nuclei and antinuclei in HBBs and 
anti-HBBs respectively
could be similar to those reported  by the \textit{AMS-02} experiment. 
Of course, one can be sure only if  and when
more rigorous nucleosynthesis calculations are done.  

On the other hand, the calculations of primordial abundances of heavy elements
like iron inside HBBs seem to be more reliable because these nuclei are the most strongly bound.
Therefore, some anti-HBB stars can be naturally poor in anti-hydrogen, and thus prone to explode as 
anti-type~Ia supernovae if they are in binary systems. This interesting issue, however, is beyond the scope of the present paper and deserves further investigations.

\section{Antinuclei in cosmic rays}
\label{s:AMS}
The \textit{Alpha Magnetic Spectrometer aboard the International Space Station} experiment \cite{AMSreview21} is currently the most sensitive experiment to detect antimatter in low-energy cosmic rays (CR). The \textit{AMS-02} collaboration reported detection of a few candidate events of antihelium nuclei with rigidity below 50~GV. \textit{AMS-02} has collected by now about $3.4\times 10^6$ positrons,  $8\times 10^5$ antiprotons and found a few antihelium candidate events (to be compared with about $N_\mathrm{He}\approx 1.3\times 10^9$ of He nuclei detected, see COSPAR-22). In so far as  the number of the detected anthelium nuclei has not been officially confirmed, in our estimate we will normalize their fraction by factor $f_{-9} \equiv 10^9 N_{\rm \overline{He}}/N_{\rm He}$ of the detected He nuclei. 
The purported dozen anti-He nuclei, if confirmed, would correspond to $f_{-9}\sim 10$. 

The spectra of CR positrons measured by the \textit{AMS-02} experiment  
are inconsistent with the hypothesis of the dominant contribution of the secondary positrons produced in the collisions of CR nuclei with the interstellar medium. A number of possible sources of CR positrons associated with high energy processes in supernova remnants, compact relativistic stars as well as the dark matter particle  decay/annihilation have been discussed,  see e.g. \cite{AMS_19,AMSreview21,DMpositr14,p_gamma14,HAWC17,0437,pulsar_halo22}. 
The origin of the CR antiprotons detected by \textit{AMS-02} \cite{AMS_pbar16} is under debate as well \cite{2015JCAP...09..023G,pbar_disc22}. In this context it is worth  discussing the possible implications of the antihelium CR flux  measurements if the candidate events will be confirmed by improved measurements.   

There are models for the high-energy antihelium nuclei production in the inelastic CR collisions  and due to dark matter (DM) annihilations (see e.g. \cite{DM2017arXiv170906507B,DM2017PhRvD..96j3021B,DM2017ICRC...35..270O,DM2020JCAP...08..048K,DMannihilCholis20,DM2021PhRvL.126j1101W,DM2021PhRvL.126j1101W} and the references therein).   
Recently, the authors of~\cite{AntiheliumCRPoulin19} presented semi-analytical estimates of the  antihelium-3 and antihelium-4 flux expected from the primary CR interactions with the interstellar medium. They found that  
the antihelium-3 flux is likely to be one to two orders of magnitude below the sensitivity of \textit{AMS-02} after five years of observations while that of antihelium-4 is below by five orders of magnitude.
They pointed out that the DM annihilation scenario likely produces not yet detectable antihelium  fluxes. They also concluded that a single antistar could explain \textit{AMS-02}
data and that there is no strong constraints on the presence of
such antistars in our Galaxy from gamma-ray astronomy.
The effect of Fermi-II re-acceleration of antideuteron and antihelium-3 nuclei produced by DM annihilation by MHD waves in the interstellar medium was discussed by \citep{DMannihilCholis20} who found that it can boost the expected fluxes of the antinuclei by an order of magnitude or more.  They concluded eventually that one can expect about one antihelium and a few antideuteron events in six years of \textit{AMS-02} data within the DM-annihilation model.  

While the production of antihelium-3 nuclei could be feasible with some DM-related coalescence and particle propagation scenarios without a conflict with the antiproton data, the confirmed detection of even a single antihelium-4 nucleus in the current \textit{AMS-02} sample would be not very easy to explain within the DM-production scenario (however, see \cite{arxiv.2211.00025}) . Then antistars or, more generally, antimatter objects could be considered as a possible explanation.  The most important distinction between the antiworlds and DM scenarios is the expected fluxes of CRs heavier than helium, which can be tested with the improved detectors. We discuss below the estimations of heavy antinuclei CR fluxes for the existing models of antistar composition.           

\subsection{Propagation of antinuclei in the Galaxy}

We start with discussing the CR antinuclei propagation in the Galaxy.
If the annihilation cross section on the interstellar material is of the order of or somewhat larger than $\sigma_{a}\sim 100$ mb, then a traversed grammage $X$ of an antinucleus of a few g~cm$^{-2}$ would be close to the maximal value $X_a\sim m_p/\sigma_a$. The typical (mean) grammage traversed by the galactic CRs estimated from the measured Li, Be, B fluxes is about 12  g~cm$^{-2}$ at a few GV rigidity (e.g. \cite{2007ARNPS..57..285S}), and it decreases with CR energy. The grammage accumulated by CRs within the accelerator is estimated to be about 0.2 g cm$^{-2}$ for a supernova remnant located in the interstellar medium with a number density of $\sim$ 1 cm$^{-3}$ (see e.g. \citep{CR_SNR_grammage15}), and it should be much smaller for sources in the halo.

Note that the  $^3$He/$^4$He ratio measured by \textit{AMS-02} \cite{AMS_helium_PRL19} decreases with rigidity (from about 0.17 at 1 GV). This is  different from the measured B/C and
B/O flux ratios, which have a maximum at about 4 GV, while at higher energies all of these three ratios have similar energy dependence \cite{AMS_helium_PRL19}. Since the inelastic interaction cross sections of helium CR with the interstellar matter are smaller than that of heavier CR nuclei (C, N, O), the helium CRs may probe larger Galactic  distances. Thus, given the mean Galactic disk ISM density $\sim 10^{-24}$ g~cm$^{-3}$ and the estimated mean grammage, a model of multiple Galactic sources is not excluded. Given the very low antihelium fluxes under discussion, even nearby extragalactic sources could contribute (see below Section \ref{s:extra}).

\subsection{Isotopic antihelium composition}

Among a dozen of antihelium nuclei reported by the  \textit{AMS-02}, a few can be antihelium-3 nuclei. This is extremely high ratio compared to the average solar wind value of $\sim 5\times 10^{-4}$ \cite{sf1980JGR....85.6021O} and somewhat larger than the  $^3$He/$^4$He ratio measured in the Galactic CRs \cite{AMS_helium_PRL19} where the CR propagation effects are important. If one considers a local antihelium CR source (with the antiisotopes ratio typical for Galactic sources) then a 1000-fold enhancement of the $^3{\overline{\rm He}}/^4\overline{\rm He}$ ratio can still be produced in antistars by analogy with the observed rare class of impulsive solar flares \cite{sf2013SSRv..175...53R,sf2020SSRv..216...24B}. 
It is also not excluded (and even expected) that HBB antistars could have an initially unusual chemical abundance enhanced with light isotopes and even metals up to iron peak
(see above Section \ref{s-BBN}).  
%\sb{In Sec.~\ref{s-BBN} we have heavy metals up to iron-peak.}

However, in view of comparable cross-sections of the $^3$He spallation production from $^4$He CRs ($\sim 55$~mb, e.g. \cite{1998ApJ...496..490R}) and in-flight annihilation reaction $^4\overline{\rm He}+p$ ($\sim 120$ mb, e.g. \cite{1994NCimR..17f...1B,2014PhRvC..89e4601L}), the high $^3{\overline{\rm He}}/{^4\overline{\rm He}}$ ratio can result from a distributed population of sources producing mostly $^4{\overline{\rm He}}$ taking into account the $^3{\overline{\rm He}}$ production by $^4\overline{\rm He}$  spallation during the  propagation in the Galaxy. The {\it ALICE} CERN experiment reported \cite{2023NatPh..19...61A} the inelastic interaction cross section of $^3\overline{\rm He}$ needed to study its propagation through the interstellar matter. The  $^3\overline{\rm He}$ nuclei can be produced in the Galaxy either in dark-matter annihilations or by CRs acceleration processes in antistars, and the authors in \cite{2023NatPh..19...61A} concluded  that the transparency of our Galaxy for $^3\overline{\rm He}$ CR propagation varies from 25\% to 90\% depending on the antinuclei momentum studied up to 10 GeV/c.

\section{Antistars as sources of anti-CRs}
\label{s:antiCR}

The \textit{AMS-02} results can be interpreted by the presence of antistars in the Galaxy. We consider several possible options for this scenario: anti-CRs from a single relatively close antistar, from outbursts of Galactic novae or type Ia supernovae, or from a collection of antistars in the a few kpc scale size Galactic halo or even in nearby galaxies.

\subsection{A single antistar: energy requirements}  

The differential flux of antihelium nuclei in the \textit{AMS-02} experiment strongly depends on the rigidity of He nuclei ($\sim R^{-2.7}$, e.g. \cite{AMS_helium_PRL19}). To estimate the total He CR flux above 10 GV, we can take the flux at 10 GV $\Phi_{\rm He}(10 GV)\approx 5$ [m$^2$~s~sr~GV]$^{-1}$ (Table 7 in \cite{2021PhR...894....1A}), so that the isotropic total flux is $dN_{\rm He}/dt/dA\sim 4\pi\times \Phi_{\rm He}(10\,\mathrm{GV}) \times 10 [\mathrm{GV}]\approx 0.06$~[cm$^2$~s]$^{-1}$. The energy flux is correspondingly $dE_{\rm He}/dt/dA\approx 10^{-3}$~erg cm$^{-2}$ s$^{-1}$. Taking an $f_{-9}\times 10^{-9}$  part in anti-He nuclei, we obtain the (presumably) isotropic anti-He energy flux $F_{\overline{\rm  He}}=dE_{\overline{\rm He}}/dt/dA\approx f_{-9}\times 10^{-12}$~erg cm$^{-2}$ s$^{-1}$.\footnote{Direct integration of $\Phi_\mathrm{He}$ above 10 GV in Table 7 from \cite{2021PhR...894....1A} would give a He flux $\sim 1.4$ times lower than our crude estimate. However, we do not take it into account in view of a larger uncertainty in the factor $f_{-9}$.}

Suppose there is one source of anti-He nuclei at a distance $r$ from the Earth producing an energy outburst $\Delta E_{\overline{\rm CR}}$ in a time interval $\Delta t$. If $\Delta t$ is much shorter that the CR diffusion time $t_{\rm dif}$, we can consider an instant injection of the anti-CR energy $\Delta E_{\overline{\rm CR}}$. For simplicity, assume a homogeneous case with a fiducial CR diffusion coefficient $D=3\times 10^{28}$~cm$^2$s$^{-1}$. Then the isotropic diffusion flux is $F_{\rm dif}= \Delta E_{\overline{\rm CR}} c/(4\pi Dt_{\rm dif})^{3/2}(1/e)$, where $c$ is the speed of light and the diffusion time is connected with the distance to the source as $r^2=4Dt_{\rm dif}$, hence 
$r\approx 200 [\mathrm{pc}] \sqrt{t_{\rm dif}/10^5\mathrm{yrs}}$. 

To get the \textit{AMS-02} energy flux of anti-He nuclei, the energy of a single anti-CR outburst should be $\Delta E_{\overline{\rm CR}}\approx 1.4\times 10^{43}[\mathrm{erg}] (f_{-9}F_{\overline{\rm He}}/10^{-12} \mathrm{erg\, cm^{-2}\, s^{-1}})(r/1\,\mathrm{kpc})^3$. Such a powerful outburst would be difficult to produce by a single flaring antistar located beyond a few parsecs away, but as we noted above, may originate from a more powerful antinova (antisupernova) outburst. We shall discuss below the possible role of the known powerful galactic CR sources -- novae and supernovae, assuming that the basic physics of particle acceleration by electromagnetic fields is not very different in the vicinity of flaring antistars compared to that we observe in the Milky Way.

\subsection{Antinovae as anti-CR sources}

To estimate the fluxes of anti-nuclei produced by a Galactic antistar population in the steady model we can rely on  the standard model of the CR propagation in the Galaxy inside a few kpc-size CR magnetic halo \cite{Berezinskii90,2007ARNPS..57..285S}. In this model, the steady power supply required to maintain the  CR fluxes observed on the Earth is $W_{\rm CR}\sim 10^{41}$~\ergs (see e.g. \cite{Berezinskii90}). Then by scaling the observed anti-CR flux to the bulk CR values, one may estimate that the low fluxes of $\overline{\rm He}$ at GeV energies would need a relatively modest power supply of $W_{\overline{\rm CR}}\sim f_{-9}\times 10^{32}$ \ergs\ in CRs to maintain the quasi-steady 
fluxes of anti-nuclear CRs if their sources are distributed over the Galaxy.
This power can be provided by e.g. flaring activity of multiple Galactic antistars (see Section \ref{s:distributed} below). 

Recent detection of gamma-rays in the 60 GeV to 250 GeV energy range from an outburst of the well known recurrent nova RS Ophiuchi (a symbiotic binary system with a white dwarf accreting mass from a red giant donor star) by the MAGIC telescope \cite{Nova_MAGIC22} proves that the novae outbursts can accelerate CRs well above the GeV energies (see also the  \textit{Fermi}/LAT spectra of the event in the 50 MeV -- 23 GeV range \citep{Nova_Fermi22}). In 2021, \textit{Fermi}/LAT telescope also detected a signal in the 0.1-100 GeV range from nova V1674 Her  that appeared 6~h after the outburst detection and lasted for 18~h with an integrated energy flux of $\sim 10^{-9}$~erg cm$^{-2}$ s$^{-1}$ \citep{nova_Her23} suggesting a gamma-ray luminosity of above $10^{36}$ erg~s$^{-1}$ at a distance of about 6~kpc advocated by the authors. Based on the analysis of the temporal evolution of the very high energy radiation from the outburst of RS Ophiuchi detected by the H.E.S.S. telescope \cite{Nova_RS_HESS22}, the authors concluded that the radiation is likely of hadronic origin and that shocks in dense winds of novae can be efficient proton accelerators up to TeV energies. While the estimated contribution from novae outbursts to the observed CR energy density was estimated in  \cite{Nova_RS_HESS22} to be subdominant compared to that from supernovae,  antinovae can still be the dominant sources of antinuclei CRs given the much lower fluxes of the antihelium CRs discussed here.          

Using the CR energy from the RS Ophiuchi outburst estimated in \cite{Nova_MAGIC22} to be about 4$\times 10^{43}$ ergs, one can conclude that just a single outburst of one Galactic antinova star of this type in a $\sim 10^5/f_{-9}$  years would be enough to maintain the steady anti-CR fluxes in the Galaxy. The Galactic novae catalog \cite{novae_distrib_MN22} accounted for the distances and properties of 402 objects, and the estimated average Galactic nova rate is 50$^{+31}_{-23}$ yr$^{-1}$ \cite{nova_rate17}.  The nova rate is broadly consistent with 28$^{+5}_{-4}$ yr$^{-1}$ obtained recently by scaling of the M31 nova rate data to the Milky Way \citep{nova_rate22}. 
New results for 
Galactic nova rate of $47.9^{+3.1}_{-8.3}$\,yr$^{-1}$ in \cite{2023arXiv230308795Z} agree with \cite{nova_rate17} but have a much higher accuracy. 

The recurrent nova RS Oph currently demonstrates eruptions every 9-27 years \cite{novae_distrib_MN22}. These numbers indicate that the putative antinova events distributed over the Galaxy with an outburst 
rate of about  10$^{-5}/f_{-9}$  of that   of RS Oph could supply quasi-
steady fluxes of the GeV antinuclei. The large power of the nova eruptions and the recurrence as high as that observed now in RS Oph  may provide the GeV anti-He flux $F_{\overline{\rm He}} \sim f_{-9}\times 10^{-12}$~erg cm$^{-2}$ s$^{-1}$ even from a single Galactic anti-nova at a few kpc distance assuming a few percent efficiency of (anti)CR acceleration. Further gamma-ray observations can constrain the number of such sources currently given by the RS Oph detection.            

While the energetics and the event rate estimations seem to be appropriate within the antinovae scenario, the key question is what kind of matter in an antinova shock can be accelerated to GeV energies? 

The fast  (possibly multiple) shocks driven by the ejected material with a complex structure, suggested by multiwavelength observations of novae  \cite{Bode_08,novae_2021ARA&A..59..391C} and involved to accelerate particles to GeV energies, are propagating through the circumstellar matter, and therefore the CR composition would reflect the circumstellar abundances. The classical nova eruption events are produced due to accretion of hydrogen-rich matter onto a white dwarf in a close binary system \cite{Bode_08,novae_2021ARA&A..59..391C}, and thus the circumstellar material in the putative antinova should be composed mainly of the binary companions material and not of the Galactic interstellar matter. Indeed, from the analysis of the hard X-ray emission detected by NuSTAR in V1674 Her \cite{nova_Her23} the authors concluded that the shock responsible for the emission was likely propagating within the nova ejecta. The shock in this case can be formed at the interaction region of the slowly moving common envelope and fast radiation-driven wind from the hot nuclear-burning white dwarf \cite{nova_Her23} and therefore propagate into the ejected envelope. Fast shocks with a velocity of $\gsim 1,000 \kms$ would accelerate the antimatter-dominated CRs to GeV energies. Simultaneous gamma-ray and optical observations of nova V906 Carinae \cite{shock_nova20} provided direct evidence for the shock-powered emission.  Modeling of gamma-ray emission generated by the internal shocks in classical novae eruptions presented in \cite{reverse_shock_novae18} offers a plausible scenario.

\subsection{CRs from type Ia anti-SN}
\label{s:aSNIa}

Yet another possibility could be an explosion of type Ia anti-SN from the old population of the Galactic halo antistars. The rate of such an exotic SNIa is highly uncertain, but most likely they occur in the Galactic halo (similar to the recently found SRG/eROSITA X-ray SNR candidate G116.6-26.1 \cite{2021MNRAS.507..971C}). Such an event could produce a flare of antinuclei CRs, including iron-group antinuclei. The available data suggest that  G116.6-26.1  is a likely type Ia halo radio-dim SNR which can be identified in sensitive X-ray surveys. The remnants of anti-SNe can be among the type Ia halo SNRs in the low-density environment which reduces the production of gamma-rays and radio emission from the forward shock, but GeV  anti-CRs accelerated in the reverse shock can escape and contribute to the Galactic anti-CRs flux. The halos of galaxies with anti-SN of type Ia  can be very extended (larger than 100 kpc) which we shall discuss below in Section \ref{s:extra}. 

We do not consider here core-collapsing SNe of types Ib/c and II, since in our scenario the antimatter massive antistars, which could produce them, have ended their life-cycle long
ago. They do not have abundant circumstellar antimatter to produce young massive antistars as is usual for ordinary stars unless in a scenario with entire antigalaxies.
SNe~Ia are believed to be produced by old low-mass binary stars (see e.g.~\cite{2018SSRv..214...72R}), most likely double white-dwarfs (the so-called double degenerate scenario e.g.~\cite{1984ApJ...284..719I,1984ApJ...277..355W,DD_type1a12}).

Supernovae (SNe) and supernova remnants (SNRs) are observed only at very low galactic z-distances in galaxies similar to Milky Way.
Pavlyuk and Tsvetkov~\cite{2016AstL...42..495P}
find that for exponential distribution
\(
\exp(-{z}/{z_0})
\)
the value of $z_0$ is $ 0.55 \pm 0.11$~kpc for SNe~Ia.
Similar values are obtained for SNRs~\cite{2015ApJ...812...37F}.
Ordinary SNRs have a $z$-distance less than the 0.5~kpc scale-height above the Milky Way plane.
They need a rather dense CSM to be observed and the local H~I halo including the Local Spur is close to the Galactic plane \cite{1986ApJ...301..380L,2009ApJ...699..716C}.
%(Lockman et al. 1986; Cersosimo et al. 2009). 
The $z$-distance of the most of metal-poor stars is also low \cite{2012ApJ...751..131B}.
%(Bovy et al. 2012).
In our scenario, antistars are located mostly in a DM halo at distances of many kpc. This means that the density of circumstellar matter around them is unusually low, and the implied mechanical braking of the blast-wave is strongly reduced in comparison with ordinary SNRs. %\sb{And strongly modified by the annihilation, but this is a separate subject of study suggested by A.Yudin within AD grant, and we should omit its discussion here.}  

Modeling of the interaction of ordinary type Ia SNe with their surrounding matter (see e.g. \cite{Dwarkadas_Chevalier98}) in addition to the blast wave demonstrated the presence of a
reverse shock which is moving into the metal-rich ejecta.  
CRs are likely accelerated in supernova remnants both in the interaction of the external shock with the circumstellar matter and by the reverse shock traveling through the metal-rich ejecta (see e.g. \cite{CR_SN_SSRv18,Raymond_ejecta_18}). 
Low surrounding density thus reduces the strength of the reverse shock in anti-SNRs.

Despite some issues with the reduced kinetic energy and low magnetic field, the reverse shock in the expanded clumpy ejecta composed from antinuclei in an anti-SN can accelerate antinuclei to relativistic energies \cite{Helder12}. Even a few percent energy efficiency of the CR acceleration by the reverse shock in an anti-SN can still power up to $\Delta E\sim 10^{48}$~ergs in anti-CRs. It should be noted here that while the reverse shocks are likely capable of producing a large amount of GeV  CRs  via the diffusive Fermi acceleration mechanism, it is still uncertain how many particles could escape from the accelerator without experiencing strong adiabatic losses in the expanding SNR. The escaping fraction should depend on the SN environment \cite{2011MNRAS.415.1807D,2011ApJ...731...87E}.     

The chemical composition of heavy antinuclei in this case can be the same as in the explosive nucleosynthesis yields, i.e. dominated by the iron-group elements \cite{1997NuPhA.621..467N,Mori_2018}.  In the likely double-degenerate scenario, the SNIa explosions are produced due to merging of two white dwarfs with the total mass exceeding the Chandrasekhar limit~\cite{1984ApJ...284..719I,1984ApJ...277..355W}. Such anti-white dwarfs can be remnants of the halo antistars and the 
anti-SN produced can be located at high distances $\sim 10$~kpc above the Galactic disk (quite atypical for normal SNIa). Their abundance and formation rate are highly uncertain, but even one such an explosion per a CR diffusion time of several million years would be sufficient to produce the observed flux of anti-CRs.

\subsection{Spatially distributed antistars}
\label{s:distributed}

The anti-CR flux can also be estimated by assuming the existence of a Galactic population of 
less powerful sources of anti-CRs, for example chromospheric
flares from antistars with properties similar to ordinary stellar flares. Then the observed antihelium to helium CR ratio $\sim f_{-9}\times 10^{-9}$ 
can be explained by the presence of $N_*$ chromospherically active antistars in the Galaxy providing a total anti-CR power of $W_{\overline{\rm CR}}\sim f_{-9}\times 10^{32}$~erg s$^{-1}$.  Assuming that the most powerful chromospheric flares in dwarf stars with energy $\Delta E\sim 10^{36}$~erg occurring on average once per $T\sim 3 \times 10^3$~yrs \cite{2022LRSP...19....2C} can accelerate $\overline{\rm He}$ CRs to GeV energies, the number of Galactic flaring antistars with the required power would be $N_{\bar f}(>10 \mathrm{GeV})\sim f_{-9}\times 10^7 (\Delta E/10^{36}\mathrm {erg})^{-1}(T/3\times 10^3\,\mathrm{yrs})$.

A subtle point is the total number of flaring antistars capable of accelerating He CRs up to the energy $\sim 10$~GeV.  Indeed, observations suggest that the power of stellar flares changes as $dN/dP\sim P^{-1.8}$ in a wide range from $\sim 10^{33}$ to $P_{\rm max}\sim 10^{38}$ erg s$^{-1}$ \cite{2021ApJ...910...41A}. The CR nuclei energy attained by Fermi-type acceleration in MHD outflow of plasma with bulk speed $u$ in a stellar flare is $E_{\rm CR}\propto (uP/c)^{1/2}$.
The measured energy of nuclei produced in powerful solar flares with $P_\odot\sim 2\times 10^{32}$ erg s$^{-1}$  is $E_{\rm max}\sim 100$  MeV \cite{2022LRSP...19....2C}. Therefore, the number of flaring antistars with potential power to accelerate 10 GeV nuclei is $N_{\bar f}(>10 \mbox{GeV})\sim N_{\rm anti*}(P_\odot/10^4 P_\odot)^{1/2} = 0.01 N_{\rm anti*}$. Therefore, the total number of flaring DM halo antistars in the Galaxy $N_{\rm anti*}\sim f_{-9}\times 10^{9}$ should be sufficient to provide the anti-He CR flux used in the above estimate. The total mass of such an antistar population is much smaller than the Galactic dark halo mass $\sim 10^{12} M_\odot$  and 
fully consistent with existing upper limits on the mass fraction of massive halo objects derived from microlensing constraints (e.g. \cite{2000ApJ...542..281A,2013MNRAS.435.1582C}).

Cosmological antistars produced by  the  HBB mechanism \cite{1993PhRvD..47.4244D,2009NuPhB.807..229D} are expected to populate a Galactic DM halo with radius $r_h\sim 150$~kpc and move with virial velocities $v\sim 10^{-3}$ \cite{2015PhRvD..92b3516B}. Their number density should vary as $\bar n_{\rm anti*}\sim 1/r^2$. Therefore, their local number density at a distance of $d_\odot\sim 8$~kpc from the Galactic center
should be $n_{\rm anti*}\approx N_{\rm anti*}/(4\pi r_h d_\odot^2) \sim 8\times 10^{-6}(N_{\rm anti*}/{10^9})$~pc$^{-3}$. Thus, with the ordinary star local number density $n_*\sim 0.15$ pc$^{-3}$,
the fraction of flaring antistars should be 
\begin{equation}
f_{\rm anti*} = \frac{n_{\rm anti*}}{n_*} \sim 5\times 10^{-5} 
\myfrac{N_{\rm anti*}}{10^9}\myfrac{150\,\mathrm{kpc}}{r_h}\,.
\end{equation}
This estimate of high-velocity antistars is consistent with upper limits on the antistar 
fraction $f_{\rm anti*}$ derived from the analysis of gamma-ray \textit{Fermi} observations \cite{2021PhRvD.103h3016D}.  
The authors of~\cite{2021PhRvD.103h3016D} concluded that the fraction of low-velocity solar-mass antistars in the Galactic disk (to be detectable by \textit{Fermi}/LAT up to 10 kpc) should be less than $2.5\times 10^{-6}$. High-velocity Galactic halo antistars, however, are much weaker restricted by gamma-ray observations (see Fig. 5 in Ref. \cite{2021PhRvD.103h3016D}).

\subsection{Extragalactic CR antinuclei}
\label{s:extra}

From the 95\% confidence upper limit for gamma-ray emission above 100 MeV  derived from the EGRET Compton GRO observations of the Small Magellanic Cloud, Sreekumar et al. \cite{EGRET93} concluded that the bulk of the GeV CR energy density is almost certainly not metagalactic, but the Galactic in origin (see also \cite{SMC_Fermi10,2014APh....53..120B} and references therein).
On the other hand, the very low fluxes of antimatter CRs under discussion are not constrained by 
these measurements and could be of extragalactic origin like the observed ultra-high energy baryonic CRs (see e.g. \cite{2014APh....53..120B}).
  
If there is a single source of anti-CR with the luminosity
$L_{\overline{\rm CR}} \sim 10^{33} [\mathrm{erg\,s^{-1}}]$ illuminating the Milky Way from the distance $r\sim 10-100$~kpc, the anti-CR power intercepted by the Milky Way 
will be $W_{\overline{\rm CR}} \sim L_{\overline{\rm CR}} \Delta \Omega/4\pi$, where $\Delta \Omega$ is the solid angle subtended by the Milky Way from this source. Thus, for the dilution factors $\Delta\Omega/4\pi\sim 1/10$ such an external source can provide the
isotropic diffusive anti-CR flux purportedly detected by the \textit{AMS-02} experiment.

While the CR anti-nuclei accelerated in the potential Milky Way antistars are mostly confined in the magnetic CR-halo of a few kpc scale size, the      
very old antistars in galaxies could populate extended virial galactic
DM halos of hundred kpc scale (see e.g. \cite{2018PhyU...61..115D} and  references therein). Analysis of {\it Fermi LAT} 1-100 GeV energy band 
observations of the Andromeda galaxy M31 \cite{M31_halo_2019} revealed an evidence for an extended $\sim 60^{\circ}$ excess over the Milky Way foreground emission which may indicate the presence of a large gamma-ray halo around M31. The giant CR halos of scale sizes $\sim$ 200 kpc  were suggested (see e.g. \cite{M31_MW_halos2021}) to explain the apparent extended gamma-ray emission from M31 \cite{M31_halo_2019}, and they are possibly relevant to the diffuse neutrino flux detected 
in the Milky Way by the {\it IceCube} Observatory. The giant halos may also be the storage of antinuclei produced by anti-SNe type Ia in the extended virial DM halos. The power $\sim f_{-9}\times 10^{36} \ergs$ is needed to support the giant halo of anti-CRs with 
the mean isotropic antihelium flux 
$F_{\overline{\rm He}} \sim f_{-9}\times 10^{-12}$~erg~cm$^{-2}$~s$^{-1}$ outside the CR sources. It can be supplied, for example, by a type Ia anti-SN rate of about $3\times 10^{-8}f_{-9}$ per year (see above Section \ref{s:aSNIa}). 

It should be noted that the flux estimations of CRs anti-nuclei accelerated in the Galactic sources discussed in Section \ref{s:distributed} and above are based on the standard model of the Galactic CRs propagation (see e.g. \cite{2007ARNPS..57..285S}). 
However, unlike the Galactic sources, the possible contribution from the sources in the giant DM halo (or from any extragalactic source) to the observed flux of GeV anti-CRs is very uncertain and could be well below the mean antihelium DM halo flux obtained in the previous paragraph. Indeed, the transport of anti-CRs produced in the circumgalactic medium can be strongly affected by the Galactic outflows driven by the either stellar feedback (e.g. \cite{wind_feedback_Hopkins12}) or by CR pressure gradients (e.g. \cite{CR_winds08,CR_winds_Crocker21,Eddington_CRlimit22}). Free escape of Galactic CRs into circumgalactic medium may produce CR currents which can amplify magnetic field to a magnitude about 2$\times 10^{-8}$ G in the Galactic wind on scales $\sim$ 10 kpc around our Galaxy \cite{Blasi_wind19}. The CR-driven outflow with the  velocity $\sim$ 10 - 100 $\kms$ which carry away the Galactic CRs would likely prevent the major fraction of GeV extragalactic anti-CRs (e.g. produced in the giant DM halo) to reach the Earth observer. The effect is similar to the heliospheric modulation of low-energy Galactic CRs (see \cite{Rankin22} for a recent review). On the other hand, there are  ultraviolet absorption-line observations indicating the presence of both matter outflows and inflows in the circumgalactic medium surrounding the Milky Way (see e.g. \cite{Richter17,in_out_flows19}), and the simulated structure of the CR-driven wind flows in a rotating galaxy has a rather complex shape \cite{CR_wind_Zir_96}. While some fraction of anti-CRs produced and stored in the giant DM halo could be carried into the Galaxy with the matter accretion inflows, the modulation of low-energy extragalactic CRs by Galactic outflows would certainly seriously increase the required power of potential extragalactic sources of anti-helium CRs given in Table \ref{t:sum}.

\section{Conclusions}
\label{s:conclusion}

\begin{table}[]
    \centering
    \begin{tabular}{lcccc}
    \hline       
    Source &\multicolumn{2}{c}{Ordinary} & \multicolumn{2}{c}{Antistar}\\
   \cline{2-5}
    & Galactic rate& $\Delta E_{CR}$ &  Required rate & Number/(distance)\\
    &yr$^{-1}$& erg &yr$^{-1}$& \\
    \hline
    Nova eruption & $\sim 50$ &  $\sim  10^{43}$  & $\sim 10^{-4}f_{-9}$  & a few \\
    SNIa    &$\sim 1/300$   &   $\lesssim 10^{48}$& $\sim 3\times 10^{-8}f_{-9}$ &  ($\sim 10$ kpc)\\
    Stellar flares & $\sim 3\times 10^4$  & $\sim 10^{36}$ & $\sim 3\times 10^{3}$  & $\sim 10^9f_{-9}$\\
    \hline
    \hline
    Extragalactic source& \multicolumn{3}{c}{$L_{\overline{\rm CR}}\sim 10^{33} f_{-9}$ erg s$^{-1}$} & $\sim 10-100$ kpc\\
    Giant halo around M31& \multicolumn{3}{c}{$L_{\overline{\rm CR}}\sim 10^{36}f_{-9}$ erg s$^{-1}$} & $\sim 700$ kpc\\ 
    \end{tabular}
    \caption{Possible sources of anti-He CRs that could provide the  \textit{AMS-02} diffusive energy flux $F_{\overline{ He}}\approx f_{-9}\times 10^{-12}$~[erg cm$^{-2}$ s$^{-1}$]. The GeV anti-helium flux from extragalactic sources is subject to a potentially large (but highly uncertain) depression factor due to the Galactic wind modulation effects. The power for extragalactic sources was not corrected for the depression factor and thus represents a lower limit for the source power required to provide the flux $F_{\overline{ He}}$ mentioned above.} 
    \label{t:sum}
\end{table}

We have discussed the possible diffusive fluxes of antihelium nuclei that can be expected from CR sources associated with antistars. The expected fluxes of both $^3\overline{\rm He}$ and $^4\overline{\rm He}$ due to CR spallation (as well as from dark matter annihilations) were estimated by \cite{AntiheliumCRPoulin19} to be below the sensitivity of
\textit{AMS-02} after five years of observations. The reported \textit{AMS-02} events, if confirmed, can be originated from anti-matter-dominated regions. A possible presence of the antinuclei CR component discussed above does not contradict
the anti-proton and positron fluxes detected in GeV range (e.g. \cite{AMS_pbar16,2015JCAP...09..023G}). These fluxes are much higher than antinuclei could contribute and are by far dominated  by the secondary products from CRs interactions and other sources.  

 The isotopic ratio of  $^3\overline{\rm He}$ and $^4\overline{\rm He}$ fluxes in the antihelium CR component can strongly constrain  possible sources of the antinuclei. Models with the distributed $^4\overline{{\rm He}}$-dominated antinuclei sources in the Galaxy (or at larger scales) would provide the isotopic ratio similar to that measured in the Galactic CRs where the $^3$He/$^4$He flux ratio is about 0.15 at $\sim$ 10 GeV/n energy \cite{AMS_helium_PRL19}. 
Therefore, a sizeable differences in the detected  $^3\overline{\rm He}$/$^4\overline{\rm He}$ flux ratio (either much smaller or larger than 0.15) would indicate either a local source of antinuclei (if the ratio is much smaller) or the $^3\overline{\rm He}$-dominated composition of antinuclei in the sources (if the ratio is much larger). 

Note that the mechanism of the Big Bang nucleosynthesis inside slowly expanding antimatter bubbles with a high baryonic density also predicts the possible formation of
antimetals, in particular the elements of 
anti-iron group. Their secondary production through 
cosmic ray collisions is absolutely excluded. The study of antimetal-rich CRs that could be produced in HBBs requires also modeling of their condensation into dust particles which is likely important for the modeling of CR acceleration (see e.g. \cite{1997ApJ...487..197E,2021MNRAS.508.1321T}).    

We have considered several possible sources of anti-CRs including chromospheric flares from halo antistars, antinova outbursts and anti-SN Ia explosions. We have found that the abundance of these sources required to explain the possible \textit{AMS-02} anti-He events  does not contradict the existing constraints on  antistars from gamma-ray measurements (see Table \ref{t:sum}).
Future measurements of CR heavy antinuclei by \textit{AMS-02} or larger-area experiments are crucial to test the exciting possibility of the presence of antistars in our Galaxy.

\section*{Acknowledgement}
We thank the anonymous referee for constructive comments which helped to improve the paper.  
AEB and ADD acknowledge the support from the RSF Grant  No. 22-12-00103 (anti-BBN, cross-sections of antinuclei inelastic interactions), and SIB thanks the RSF Grant No. 19-02-00229 (formation of peculiar SNe in the early Universe).  

%\bibliographystyle{JHEP}
%\bibliography{antistars}
% if your bibtex file is called example.bib

\providecommand{\href}[2]{#2}\begingroup\raggedright\endgroup

\end{document}